\documentclass[smallextended]{svjour3}

\usepackage[T1]{fontenc}

\usepackage{url}
\usepackage[misc,geometry]{ifsym}
\usepackage{float}
\usepackage{hyperref}
\usepackage{algorithm}
\usepackage{algpseudocodex}
\usepackage{amsmath}
\usepackage{array}
\usepackage{makecell}

\usepackage{listings}     
\usepackage{lstautogobble}  
\usepackage{color}          
\usepackage{inconsolata}    

\definecolor{bluekeywords}{rgb}{0.13, 0.13, 1}
\definecolor{greencomments}{rgb}{0, 0.5, 0}
\definecolor{redstrings}{rgb}{0.9, 0, 0}
\definecolor{graynumbers}{rgb}{0.5, 0.5, 0.5}

\usepackage{listings}
\begin{document}

\title{PyOptInterface: Design and implementation of an efficient modeling language for mathematical optimization}

\titlerunning{PyOptInterface: Design and implementation}

\author{Yue Yang \and Chenhui Lin \and Luo Xu \and Wenchuan Wu}
\authorrunning{Y.~Yang, C.~Lin, L.~Xu, W.~Wu}

\institute{
	Yue Yang \quad \Letter
	\at Hefei University of Technology, Hefei, China
	\\\email{yue.yang@hfut.edu.cn}
	\and
	Chenhui Lin, Wenchuan Wu
	\at Tsinghua University, Beijing, China
	\\\email{linchenhui@tsinghua.edu.cn, wuwench@tsinghua.edu.cn}
	\and
	Luo Xu
	\at Princeton University, Princeton, USA
	\\\email{luoxu@princeton.edu}
}

\date{Received: date / Accepted: date}

\maketitle

\begin{abstract}
	This paper introduces the design and implementation of PyOptInterface, a modeling language for mathematical optimization embedded in Python programming language. PyOptInterface uses lightweight and compact data structure to bridge high-level entities in optimization models like variables and constraints to internal indices of optimizers efficiently. It supports a variety of optimization solvers and a range of common problem classes. We provide benchmarks to exhibit the competitive performance of PyOptInterface compared with other state-of-the-art modeling languages.
\end{abstract}

\section{Introduction}

PyOptInterface is a modeling language for mathematical optimization embedded in Python programming language. It is designed as a high-level abstraction layer on low-level interfaces of optimizers to facilitate the formulation of optimization models in an optimizer-agnostic manner. PyOptInterface supports a variety of optimization solvers and a range of common problem classes, including linear programming, quadratic programming, mixed integer programming and second-order cone programming. By utilizing an enhanced data structure to map high-level entities (variables and constraints) to internal indices of optimizers with compact memory layout, PyOptInterface remains lightweight and efficient when handling large-scale optimization models.

The paper is organized as follows: Section \ref{sec:design} describes the design of efficient mapping from high-level entities to internal indices of optimizers. Section \ref{sec:implementation} introduces the implementation of PyOptInterface. Section \ref{sec:benchmarks} provides benchmarks to compare the performance of PyOptInterface with other modeling languages.

\section{Design of efficient mapping from entities to indices}\label{sec:design}

As an introductory example, we use a code snippet of PyOptInterface to declare three nonnegative variables $x,y,z$ and construct a linear constraint $x+2y+3z \leq 1$ with Gurobi as the optimizer.

\begin{lstlisting}[language = Python]
	import pyoptinterface as poi
	from pyoptinterface import gurobi
	
	model = gurobi.Model()
	x = model.add_variable(lb=0.0, name="x")
	y = model.add_variable(lb=0.0, name="y")
	z = model.add_variable(lb=0.0, name="z")
	constraint = model.add_linear_constraint(x+2*y+3*z, poi.Leq, 1.0)
\end{lstlisting}

The code above is roughly equivalent to the following C code that uses the low-level interface of Gurobi directly.

\begin{lstlisting}[language = C]
	GRBmodel* model;
	// omit the initialization of model
	GRBaddvar(model, 0, NULL, NULL, 0.0, 0.0, GRB_INFINITY, GRB_CONTINUOUS, "x");
	GRBaddvar(model, 0, NULL, NULL, 0.0, 0.0, GRB_INFINITY, GRB_CONTINUOUS, "y");
	GRBaddvar(model, 0, NULL, NULL, 0.0, 0.0, GRB_INFINITY, GRB_CONTINUOUS, "z");
	
	int cind[] = {0, 1, 2};
	double cval[] = {1.0, 2.0, 3.0};
	GRBaddconstr(model, 3, cind, cval, GRB_LESS_EQUAL, 1.0, "constraint");
\end{lstlisting}

Comparing the two approaches, lower-level interface of optimizers is only aware of numeric column indices of variable and row indices of constraint, and users must manage these indices manually, which is error-prone and not flexible. By contrast, the modeling language of mathematical optimization provides high-level entities to represent the variables and constraints in optimization model. Users can define and manipulate these high-level entities as persistent objects and use them to formulate the optimization model in a structured way.

Thus, it is the crucial responsibility of modeling language to construct an efficient mapping from user-defined entity (variables and constraints) to internal column and row indices understood by optimizers. We rely on the mapping to create, modify and delete these high-level entities by invoking low-level index-based interface of optimizers correspondingly.

If we add $N$ variables to an optimization model, their column indices from the perspective of optimizer are a continuous and monotonic series from $0$ to $N-1$. A natural idea to design the mapping is to allocate an array with length $N$ filled from 0 to $N-1$. Each entity of variable stores an index to the array and their column index is the corresponding element of array. With the array-based mapping, obtaining the index of an entity and adding new entities has $O(1)$ complexity which are very efficient. The array-based mapping has been widely used by higher-level interface of optimizers, such as Python binding of HiGHS\cite{huangfu_2018}.

The memory consumption of array-based mapping depends on the size of index stored in array. If the type of internal column index used by optimizer is \texttt{int}, then each entity occupies 4 bytes (32 bits) of memory.

The array-based mapping is not efficient enough to handle removal of entities. When an entity with index $i$ is removed from the mapping, the array-based mapping must be updated in 2 steps:

\begin{enumerate}
	\item	Mark the element at index $i$ is invalid
	\item	Subtract all elements after index $i$ by 1
\end{enumerate}

Deleting one entity from array-based mapping has $O(N)$ complexity because the worst case happens when the first entity is deleted and the indices of all subsequent entities need to be updated.
As a more general case, deleting $M$ entities from the mapping has $O(MN)$ complexity.
There is an optimized deletion strategy to reduce the complexity from $O(MN)$ to $O(M\log M + N\log M)$ used by several solver wrappers in JuMP.jl\cite{lubin2023jump} ecosystem.
\begin{enumerate}
	\item Sort the $M$ entities to delete as a list. ($O(M\log M)$)
	\item For each entity in the array, use binary search ($O(\log M)$) to find how many deleted entities are in front of it and denote it as $m$, then subtract its index by $m$.
\end{enumerate}

We notice that an entity in an optimization model only has two states that can be represented by single bit: 1 represents that the entity is added to the model, 0 represents that the entity is deleted from the model. Thus, we can use a bitmap to represent the states of entities. For example, if there are 5 variables in a model, the bitmap of variables are [1, 1, 1, 1, 1]. The bitmap becomes [1, 1, 0, 1, 1] after the third variable is deleted.

If we follow the convention of C language and count the internal index used by optimizer from 0, the internal index of entity represented by the $i^{th}$ bit is defined as follows:

\begin{equation}
	I(i) =
	\begin{cases}
		0                                              & \text{if } i = 0 \text{ and } \text{bitmap}[0] = 1, \\
		\displaystyle\sum_{k=0}^{i-1} \text{bitmap}[k] & \text{if } \text{bitmap}[i] = 1,                    \\
		\text{INVALID\_INDEX}                          & \text{if } \text{bitmap}[i] = 0.
	\end{cases}
\end{equation}

There are two trivial ways to compute $I(i)$.
\begin{enumerate}
	\item Always scan the bitmap from the beginning, it has $O(N)$ complexity.
	\item Precompute $I(i)$ from $i=0$ to $i=N-1$. It is essentially the same as array-based mapping.
\end{enumerate}

In fact, $I(i)$ is closely related to the \texttt{rank} query of bitmap, so a data structure called BItmap with Progressively Updated Ranks (BIPUR) is proposed to compute $I(i)$ efficiently as an improved version of the second approach. The improvement of BIPUR is based on two intuitive observations:
\begin{enumerate}
	\item The bits can be separated into chunks and $I(i)$ is precalculated for each chunk instead of each bit to save memory.
	\item Calculating $I(i)$ only requires information about bits in front of the $i^{th}$ bit rather than the whole bitmap, so we defer the computation until it is necessary. The precomputed $I(i)$ are cached and reused for subsequent bits to save computational time.
\end{enumerate}

The BIPUR data structure includes the following components:
\begin{enumerate}
	\item The bits are separated into contiguous 64-bit chunks and stored in a \texttt{std::vector<uint64\_t>} named \texttt{chunks}
	\item For each chunk, it maintains two kind of caches: \texttt{cumulated\_ranks} and \texttt{chunk\_ranks}. For the $i^{th}$ chunk, \texttt{cumulated\_ranks[i]} records the number of $1$ in all previous chunks \texttt{chunks[0, i)}, and \texttt{chunk\_ranks[i]} records the number of $1$ in the $i^{th}$ chunk.
	\item A counter \texttt{next\_bit} points to the next bit.
	\item An index \texttt{last\_correct\_chunk} indicates the progress of precomputed $I(i)$, ensuring that \texttt{cumulated\_ranks[0, last\_correct\_chunk]} and \texttt{chunk\_ranks[0, last\_correct\_chunk)} are correct.
\end{enumerate}

\begin{algorithm}
	\caption{Operations of BIPUR}\label{alg:bipur}
	\begin{algorithmic}[1]

		\Procedure{initialize}{}
		\State $chunks \gets [0]$
		\LComment{The chunks are contiguous 64-bit integers used as bitmap}
		\State $cumulated\_ranks \gets [0]$
		\State $chunk\_ranks \gets [-1]$
		\LComment{The number of $1$ in current chunk is uncounted}
		\State $next\_bit \gets 0$
		\State $last\_correct\_chunk \gets 0$
		\EndProcedure

		\Procedure{add\_entity}{}
		\If{$next\_bit \leq \text{length}(chunks) \times 64$}
		\State $bitmap[next\_bit] \gets 1$
		\Else
		\State $chunks.\text{append}(1)$
		\State $cumulated\_ranks.\text{append}(0)$
		\State $chunk\_ranks.\text{append}(-1)$
		\EndIf
		\State $bit\_location \gets next\_bit$
		\State $next\_bit \gets next\_bit + 1$
		\State \Return $bit\_location$
		\LComment{Handle of the new entity}
		\EndProcedure

		\Procedure{delete\_entity}{$bit\_location$}
		\If{$bitmap[bit\_location] == 0$}
		\Comment{The entity has already been deleted}
		\State \Return
		\EndIf
		\State $bitmap[bit\_location] \gets 0$
		\State $(chunk\_index, bit\_index) \gets \texttt{divmod}(bit\_location, 64)$
		\State $chunk\_ranks[chunk\_index] \gets -1$
		\LComment{The number of $1$ in current chunk should be re-counted}
		\If{$last\_correct\_chunk > chunk\_index$}
		\State $last\_correct\_chunk \gets chunk\_index$
		\EndIf
		\EndProcedure

		\Procedure{calculate\_index}{$bit\_location$}
		\If{$bitmap[bit\_location] == 0$}
		\Comment{The entity has already been deleted}
		\State \Return $\text{INVALID\_INDEX}$
		\EndIf
		\State $(chunk\_index, bit\_index) \gets \texttt{divmod}(bit\_location, 64)$
		\If{$last\_correct\_chunk < chunk\_index$}
		\State precompute\_chunk\_rank($chunk\_index$)
		\EndIf
		\State $current\_chunk \gets chunks[chunk\_index]$
		\State $chunk\_bits \gets \text{the number of $1$ in} \: current\_chunk[0, bit\_index)$
		\LComment{Can be accelerated by \texttt{popcount} instruction}
		\State \Return $cumulated\_ranks[chunk\_index] + chunk\_bits$
		\EndProcedure

		\Procedure{precompute\_chunk\_rank}{$chunk\_index$}
		\For{$j \gets last\_correct\_chunk, chunk\_index-1$}
		\If{$chunk\_ranks[j] == -1$}
		\State $chunk\_ranks[j] \gets $ the number of $1$ in $chunks[j]$ \LComment{Can be accelerated by \texttt{popcount} instruction}
		\EndIf
		\State $cumulated\_ranks[j + 1] \gets cumulated\_ranks[j] + chunk\_ranks[j]$
		\EndFor
		\State $last\_correct\_chunk \gets chunk\_index$
		\EndProcedure
	\end{algorithmic}
\end{algorithm}

Five key operations of BIPUR are described in Algorithm \ref{alg:bipur}.
\begin{itemize}
	\item The \texttt{initialize} operation initializes the BIPUR data structure.
	\item The \texttt{add\_entity} operation adds a new entity to the bitmap and returns the location of bit as the handle of new entity.
	\item The \texttt{delete\_entity} operation deletes an entity from the bitmap.
	\item The \texttt{calculate\_index} operation calculates $I(i)$ of the entity corresponding to a specific bit.
	\item The \texttt{precompute\_chunk\_rank} operation is called to precompute number of $1$ in front of a specific chunk and caches the result.
\end{itemize}

When BIPUR is used as the mapping from high level entity (variables and constraints) to column or row index of optimizer, the operational complexity can be analyzed as follows:
\begin{enumerate}
	\item Adding an entity (create a new variable or constraint): The \texttt{add\_entity} operation appends 1 bit to the bitmap and has $O(1)$ complexity.
	\item Deleting an entity (remove a variable or constraint from model): The \texttt{delete\_entity} operation flips a bit from 1 to 0 and set flags to indicate that $I(i)$ after the deleted bit need to be re-calculated, so its complexity is $O(1)$.
	\item Deleting $M$ entities in a batch: we only need to repeat the deletion procedure for $M$ times, so the complexity is $O(M)$.
	\item Calculating the internal index of entity at the $i^{th}$ bit (query the index of a specific variable or constraint): The operation is divided into 2 steps:
	      \begin{enumerate}
		      \item The first step precomputes the number of $1$ in all previous chunks and caches the result. The complexity is proportional to the number of chunks to traverse.
		      \item The second step computes the number of $1$ in front of the the $i^{th}$ bit in current 64-bit chunk.
	      \end{enumerate}
	      In both steps, we can use the hardware-accelerated \texttt{popcount} instruction of modern processors to count the number of $1$ in a 64-bit chunk within constant time. Thus, the complexity of the second step is $O(1)$. The complexity of the first step is proportional to the number of chunks to traverse.

	      If we calculate the index of all entities from $0$ to $N-1$, the first step is executed $N/64$ times to propagate the precomputed number of $1$ by one chunk, and the second step is executed $N$ times, so the amortized complexity is $O(1)$. After that, calculating $I(i)$ for any entity only needs to execute the second step, so the complexity is still $O(1)$.
\end{enumerate}

Supposing the internal index used by optimizer is $B$ bits ($B=32$ for \texttt{int}) and the number of entities is $N$, the memory consumption of BIPUR is analyzed as follows:
\begin{enumerate}
	\item The bitmap of $N$ entities needs $N$ bits.
	\item The number of chunks is $N/64$, each element in \texttt{cumulated\_ranks} occupies $B$ bits, and each element in \texttt{chunk\_ranks} occupies $8$ bits (because the number of $1$ in a 64-bit chunk is at most 64 and can be represented in 8 bits).
\end{enumerate}

The average memory consumption per entity of BIPUR is $(N + N/64 \times (B+8)) / N = 1 + (B+8) / 64$ bits. In the array-based mapping, the memory consumption is $B$ bits per entity. Thus, the memory layout of BIPUR is much smaller than the array-based mapping.

\begin{table}[!ht]
	\centering
	\begin{tabular}{l | c | c }
		Operation                    & Array-based mapping    & BIPUR              \\
		\hline
		Add a new entity             & $O(1)$                 & $O(1)$             \\
		Delete an existing entity    & $O(N)$                 & $O(1)$             \\
		Delete $M$ existing entities & $O(M\log M + N\log M)$ & $O(M)$             \\
		Calculate index of entity    & $O(1)$                 & $O(1)$ (amortized)
	\end{tabular}
	\caption{Complexity of array-based mapping and the proposed BIPUR mapping.}
	\label{tab:complexity}
\end{table}

\begin{table}[!ht]
	\centering
	\begin{tabular}{c | c | c }
		       & Array-based mapping & BIPUR \\
		\hline
		$B=32$ & 32                  & 1.625 \\
		$B=64$ & 64                  & 2.125
	\end{tabular}
	\caption{Average bits required to store the mapping from one entity to its internal index when the internal index used by optimizer is $B$ bits.}
	\label{tab:memory}
\end{table}

As summarized in Table \ref{tab:complexity} and Table \ref{tab:memory}, the BIPUR mapping is more efficient than the array-based mapping in terms of both complexity and memory consumption.

\section{Implementation of efficient modeling language for multiple optimizers}\label{sec:implementation}

PyOptInterface is implemented as a thin abstraction layer on the low-level interface of optimizers to provide common high-level functionalities of modeling language. The core part is implemented in C++ to communicate with different optimizers via their C interface, and user-friendly Python interfaces are provided to extend and wrap the C++ core. The code is available at \url{https://github.com/metab0t/PyOptInterface}. The documentation including many examples is hosted at \url{https://metab0t.github.io/PyOptInterface/}.

PyOptInterface provides two kinds of entities for users: \texttt{VariableIndex} and \texttt{ConstraintIndex} classes, which are both implemented as lightweight handles pointing to bits in the BIPUR data structure. The \texttt{VariableIndex} is mapped to the column index of optimizer through the BIPUR data structure. The indices of constraints are more complex, because different types of constraints are usually numbered separately by the optimizer. For example, Gurobi has independent counters for 4 kinds of constraints: linear, quadratic, special ordered set (SOS) and general constraints. Thus, each kind of constraint has its own BIPUR mapping to avoid conflicts. In \texttt{ConstraintIndex}, the \texttt{type} field is used to discriminate which instance of BIPUR it points to, and the \texttt{bit\_location} field stores the index of bit in the corresponding BIPUR.

\begin{lstlisting}[language = C]
	struct VariableIndex {
		int bit_location;
	};
	struct ConstraintIndex {
		uint8_t type;
		int bit_location;
	};
\end{lstlisting}

Based on abstractions of variables and constraints, PyOptInterface provides a unified interface for various optimizers including Gurobi\cite{gurobi}, COPT\cite{copt}, Mosek\cite{mosek}, and HiGHS\cite{huangfu_2018} by defining a small number of standard operations to create, modify, query and delete variables and constraints, and users can manipulate the optimization model in an optimizer-agnostic manner. These operations are instantly passed to optimizer after mapping of BIPUR, so PyOptInterface only stores several BIPUR data structures for variables and constraints in memory and completely relies on the underlying optimizer to maintain the  model internally. Table~\ref{tab:translation} shows the translation table from several operations in PyOptInterface to corresponding C interfaces in Gurobi and HiGHS as an example.

\begin{table}[!ht]
	\centering
	\begin{tabular}{l | l | l }
		Operation                           & Gurobi                              & HiGHS                           \\
		\hline
		\texttt{add\_variable}              & \texttt{GRBaddvar}                  & \texttt{Highs\_addCol}          \\
		\texttt{delete\_variable}           & \texttt{GRBdelvars}                 & \texttt{Highs\_deleteColsBySet} \\
		\texttt{add\_linear\_constraint}    & \texttt{GRBaddconstr}               & \texttt{Highs\_addRow}          \\
		\texttt{add\_quadratic\_constraint} & \texttt{GRBaddqconstr}              & \emph{Not Available}            \\
		\texttt{add\_sos\_constraint}       & \texttt{GRBaddsos}                  & \emph{Not Available}            \\
		\texttt{delete\_constraint}         & \makecell[l]{\texttt{GRBdelconstrs}                                   \\\texttt{GRBdelqconstrs}\\\texttt{GRBdelsos}} & \texttt{Highs\_deleteRowsBySet}
	\end{tabular}
	\caption{Translation table from operations in PyOptInterface to C interfaces in optimizers.}
	\label{tab:translation}
\end{table}

Compared with using these low-level interfaces of optimizers directly, the extra overhead of PyOptInterface is mainly from two aspects:
\begin{enumerate}
	\item \emph{The mapping process from entity to index}: It has been optimized by the design of BIPUR data structure and has $O(1)$ complexity.
	\item \emph{The cost to store and manage entities in Python}: Each entity is essentially a handle of numeric index and only occupies a few bytes of memory, and the memory consumption brought by high-level abstraction is also reduced due to the compact memory layout of the BIPUR data structure.
\end{enumerate}

As shown in Table~\ref{tab:translation}, some constraints are not supported by all optimizers, such as SOS constraints and second-order cone constraints. MP library in AMPL\cite{fourer1995ampl} and MathOptInterface.jl\cite{legat2021mathoptinterface} in JuMP.jl all have automatic reformulation capability to rewrite these constraints into equivalent constraints that are natively supported by the optimizer. PyOptInterface provides a similar lightweight and pluggable constraint reformulation mechanism via monkeypatching. The general reformulation procedure can be implemented in Python and attached to the instance of optimizer. For example, the code snippet below shows how to make a \texttt{Model} class support second-order cone constraint by reformulating it as a quadratic constraint.

\begin{lstlisting}[language = Python]
	def bridge_soc(model, cone_variables, name=""):
		"""
		Convert a second-order cone constraint to a quadratic constraint.
		x[0] >= sqrt(x[1]^2 + ... + x[n]^2) to x[0]^2 - x[1]^2 - ... - x[n]^2 >= 0
		"""
		expr = ScalarQuadraticFunction()
		x0 = cone_variables[0]
		expr.add_quadratic_term(x0, x0, 1.0)
		for xi in cone_variables[1:]:
			expr.add_quadratic_term(xi, xi, -1.0)
		con = model.add_quadratic_constraint(expr, poi.Geq, 0.0, name)
		return con

	class Model:
		def __init__(self):
			... # omit the initialization of model
			self.add_second_order_cone_constraint = types.MethodType(bridge_soc, self)
\end{lstlisting}

\section{Benchmarks}\label{sec:benchmarks}

In order to evaluate the efficiency of modeling languages, we extend the benchmark suite in the JuMP paper \cite{lubin2023jump}. Four facility location models and four linear quadratic control problems with different sizes are selected as test cases. For each model, the total time of modeling language to generate model and pass it to the optimizer is measured with Gurobi and COPT as optimizer respectively, and the time limit of optimizer is set to 0.0 seconds to avoid the influence of solution process. All code to run the benchmarks is available at \url{https://github.com/metab0t/PyOptInterface_benchmark}.

\begin{table}[!ht]
	\centering
	\begin{tabular}{l | r | c | c | c | c | c}
		Model     & Variables & C++  & PyOptInterface & JuMP & gurobipy & Pyomo \\
		\hline
		fac-25    & 67651     & 0.2  & 0.2            & 0.2  & 1.2      & 4.1   \\
		fac-50    & 520301    & 0.8  & 1.2            & 1.8  & 9.7      & 32.7  \\
		fac-75    & 1732951   & 2.7  & 4.1            & 6.6  & 32.5     & 119.3 \\
		fac-100   & 4080601   & 6.3  & 10.0           & 17.8 & 79.1     & 286.3 \\
		          &           &      &                &      &                  \\
		lqcp-500  & 251501    & 0.9  & 1.5            & 1.3  & 6.3      & 23.8  \\
		lqcp-1000 & 1003001   & 3.7  & 6.0            & 6.1  & 26.7     & 106.6 \\
		lqcp-1500 & 2254501   & 8.3  & 14.0           & 17.7 & 61.8     & 234.0 \\
		lqcp-2000 & 4006001   & 14.5 & 24.9           & 38.3 & 106.9    & 444.1
	\end{tabular}
	\caption{Time (second) to generate model and pass it to Gurobi optimizer.}
	\label{tab:gurobi_benchmark}
\end{table}

\begin{table}[!ht]
	\centering
	\begin{tabular}{l | r | c | c | c | c | c}
		Model     & Variables & C++  & PyOptInterface & JuMP & coptpy & Pyomo \\
		\hline
		fac-25    & 67651     & 0.3  & 0.2            & 0.3  & 0.6    & 4.1   \\
		fac-50    & 520301    & 2.2  & 1.5            & 2.7  & 5.4    & 32.8  \\
		fac-75    & 1732951   & 8.1  & 6.6            & 10.2 & 20.3   & 117.4 \\
		fac-100   & 4080601   & 22.4 & 23.4           & 30.3 & 58.0   & 284.0 \\
		          &           &      &                &      &                \\
		lqcp-500  & 251501    & 3.8  & 3.1            & 3.0  & 6.6    & 26.4  \\
		lqcp-1000 & 1003001   & 16.0 & 15.5           & 13.9 & 28.1   & 112.1 \\
		lqcp-1500 & 2254501   & 37.6 & 32.4           & 33.7 & 64.6   & 249.3 \\
		lqcp-2000 & 4006001   & 68.2 & 60.3           & 66.2 & 118.4  & 502.4
	\end{tabular}
	\caption{Time (second) to generate model and pass it to COPT optimizer.}
	\label{tab:copt_benchmark}
\end{table}

In our benchmark, we include the following modeling languages:
\begin{enumerate}
	\item \texttt{C++}: Official C++ interfaces to Gurobi and COPT.
	\item \texttt{PyOptInterface}: The Python interface to Gurobi and COPT implemented by our paper.
	\item \texttt{JuMP}: An open-source algebraic modeling language in Julia, it is used with \texttt{Gurobi.jl} and \texttt{COPT.jl} to interact with optimizers. In this paper, we use the direct mode feature of JuMP to skip an additional cache of the problem. The Julia code is warmed up by running a small case before the benchmark to rule out compilation latency of Julia.
	\item \texttt{gurobipy} and \texttt{coptpy}: Official Python interfaces to Gurobi and COPT.
	\item \texttt{Pyomo}\cite{hart2017pyomo}: An open-source algebraic modeling language in Python. In this paper, we use the in-memory persistent solver interface of Pyomo to interact with Gurobi and COPT without writing the model to disk.
\end{enumerate}

The software used to run benchmark includes \texttt{PyOptInterface} v0.1.0, Python v3.11.8, Gurobi v11.0.1, \texttt{gurobipy} v11.0.1, COPT v7.1.1, \texttt{coptpy} v7.1.1, Julia v1.10.2, \texttt{JuMP.jl} v1.21.0, \texttt{Gurobi.jl} v1.2.3, \texttt{COPT.jl} v1.1.15 and Pyomo v6.7.1. The hardware to run benchmark is a Windows laptop with i7-1360P CPU and 32GB RAM.

Analyzing the results presented in Table \ref{tab:gurobi_benchmark} and Table \ref{tab:copt_benchmark}, we observe a significant variation in the time required to generate and pass models to the optimizers across different modeling languages. This variation underscores the impact of the modeling language's design and implementation on its efficiency, particularly for larger-scale optimization problems.

The \texttt{C++} interface showcases the lowest generation times in most cases because of the efficiency of using native code and direct interaction with the optimizer, but it is optimizer-specific and less flexible than high-level modeling languages.

While slower than the \texttt{C++} interface, \texttt{PyOptInterface} demonstrates competitive performance in benchmark. It consistently outperforms other Python-based modeling languages and is on par or faster than \texttt{JuMP} in most cases. The gap of performance is particularly evident for larger models where the efficiency of our lightweight and compact architecture becomes more pronounced.

The official Python interfaces \texttt{gurobipy} and \texttt{coptpy} exhibit longer generation times compared to \texttt{PyOptInterface}. This suggests that there may be inefficiencies in their design or the way they handle model construction and communication with the optimizer.

\texttt{Pyomo} consistently shows the longest generation times across all test cases. This could be attributed to its general-purpose design, which aims to support a wide range of optimizers and optimization problem types.

The benchmark results illustrate the advantage of PyOptInterface as a compelling choice for optimization tasks, balancing the need for speed with the flexibility and productivity benefits of Python. It closely matches or outperforms more established modeling languages like JuMP.jl and Pyomo in terms of performance to construct optimization models.

\bibliographystyle{spmpsci}
\bibliography{main}

\end{document}